\documentclass{emulateapj}

\newcommand{\ex}[1]{\mbox{$\times 10^{#1}$}}

\newcommand{\muas}{\mbox{$\mu$as}}

\newcommand{\kms}{\mbox{km s$^{-1}$}}

\newcommand{\Jb}{\mbox{Jy bm$^{-1}$}}
\newcommand{\Ra}[4]{\mbox{${#1}^{\rm h} \; {#2}^{\rm m} \; {#3}\fs{#4} $}}
\newcommand{\dec}[4]{\mbox{${#1}\arcdeg \; {#2}\arcmin \; {#3}\farcs{#4} $}}

\shortauthors{Bietenholz \& Bartel}
\shorttitle{Expansion Velocity of SN 2001em}
\begin{document}
      
\title{An Upper Limit on the Expansion Velocity of GRB Candidate SN
2001em}

\author{M. F. Bietenholz and N. Bartel} 
\affil{Department of Physics and Astronomy, York University, Toronto,
M3J~1P3, Ontario, Canada}

\journalinfo{{\em Accepted to the Astrophysical Journal Letters}} 
 
\begin{abstract}
We report on VLBI observations of the Type I b/c supernova 2001em,
three years after the explosion.  It had been suggested that SN~2001em
might be a jet-driven gamma ray burst (GRB), with the jet oriented far
from the line of sight so that the GRB would not be visible from
earth. To test this conjecture, we determined the size of
\objectname[]SN~2001em. It is only marginally resolved at our
resolution of $\sim$0.9~mas.
The $3\sigma$ upper limit on the major axis angular size of the radio
source was 0.59~mas (FWHM of an elliptical Gaussian), corresponding to
a one-sided apparent expansion velocity of $70,000$~\kms\ for a
distance of 80~Mpc.  No low-brightness jet was seen in our image to a
level of 4\% of the peak brightness.
If we assume instead a spherical shell geometry typical of a
supernova, we find the angular radius of SN~2001em was
$0.17_{-0.10}^{+0.06}$~mas, implying an isotropic expansion velocity
of $20,000_{-12,000}^{+7,000}$~\kms, which is comparable to the
expansion velocities of supernova shells.
Our observations, therefore, are not consistent with a
relativistically expanding radio source in SN~2001em, but are instead
consistent with a supernova shell origin of SN~2001's radio emission.
\end{abstract}

\section{INTRODUCTION}

Supernova 2001em was discovered on 2001 September 15
\citep{Papenkova+2001}, at a distance of $\sim 80$~Mpc
\citep[e.g.,][assuming $H_0$ = 70 \kms~Mpc$^{-1}$]{SoderbergGS2004,
Huchra+1992}, in the galaxy \objectname{UGC~11794}
(\objectname{NGC~7112}).  It
was classified as Type Ib/c on the basis of an early optical spectrum
\citep{FilippenkoC2001}.  A recent spectrum, however, was unusual for
a Type Ib/c supernova, being dominated by a broad H$\alpha$ line (FWHM
of 1800~\kms) suggestive of strong interaction with the circumstellar
medium \citep{SoderbergGS2004}.  SN~2001em was recently also detected
in the radio \citep{Stockdale+2004}, with an 8.4~GHz flux density of
1.15~mJy on 2003 October 17.  It is unusually radio luminous, more so
than any other observed Type Ib/c supernova at its age.  Furthermore,
SN~2001em's flux density rose to 1.48~mJy by 2004 January 31, that is
$\sim 1000$~d after the explosion, which is also unusual behavior for
a Type Ib/c SN, whose radio light curves usually peak less than 1 year
after the explosion.  In addition, SN~2001em had a radio spectral
index of $\alpha \sim -0.37 \; (S_\nu \propto \nu^\alpha)$ between 5
and 15~GHz, which is flatter than is usually seen in radio supernovae
older than one year.

The unusual characteristics of SN~2001em prompted the suggestion by
\citet{GranotR2004} that it might be associated with a gamma ray burst
(GRB)\@.  There is mounting evidence that at least some GRBs are
associated with supernovae.  In particular, two GRBs have now been
reliably associated with observed supernovae: GRB~980425 with
SN~1998bw \citep{Galama+1998} and GRB~030329 with SN~2003dh
\citep{Stanek+2003, Hjorth+2003}, both of which were of Type Ib/c.

Most current GRB models rely on the collapse of a stellar mass object
to a black hole, which powers relativistic jets \citep[see
e.g.,][]{MacFadyenW1999, ZhangM2004}.
GRB jets are thought to initially be highly relativistic and have
consequently highly beamed emission.  GRB afterglows, particularly in
the radio, are seen as the jets decelerate and their emission becomes
more isotropic. If the jet model for GRBs is correct, it follows that
only a fraction of the jets are oriented near the line of sight and
are visible from earth as GRBs.  The remainder will not produce a GRB
event visible from earth.  The more isotropic afterglow emission,
however, in particular radio emission from the jet, may be detectable.
One might therefore hope to detect a signature of a GRB in some Type
I~b/c supernovae. \citet{SoderbergFW2004} have estimated the fraction
of GRB events visible from earth as being $<6$\% \citep[see
also][]{Berger+2003, Frail+2001}.  Radio searches for GRB signatures
in Type I~b/c supernovae were carried out by \citet{SoderbergFW2004}
and \citet{Berger+2003}, but none have been detected to date.
It was also predicted that some GRB jets, especially ones not aligned
near the line of sight, might be resolvable with very-long-baseline
interferometry (VLBI)
\citep{Paczynski2001, GranotL2003}.  Recently, Taylor et al.\ (2004;
see also Taylor et al.\ 2005)
resolved GRB~030329 using VLBI observations, and showed that the radio
source was expanding at an average apparent velocity of $\sim 4\,c$,
consistent with the expectations.

\citet{GranotR2004} showed that the characteristics of SN~2001em's
radio emission, in particular the delayed rise in flux density coupled
with a radio spectral index of $\alpha \sim -0.4$, are difficult to
reconcile with the circumstellar interaction model of supernova radio
emission, but are consistent with the expectations for an off-axis
relativistic jet.  Such a jet would be expected to become
non-relativistic after a few years, with the radio emission reaching
its maximum near the time of this transition \citep{GranotS2002,
GranotL2003}.  This behavior is consistent with the radio lightcurve
of SN~2001em.  The recent rise in X-ray flux from SN~2001em
\citep{PooleyL2004} is also consistent with the possibility that
SN~2001em was an off-axis GRB\@.  Could SN~2001em harbor a GRB jet
transverse to the line of sight?

At 80~Mpc, the length of a transverse jet which had expanded
relativistically since the explosion date of SN~2001em ($\sim$ 2001
September 15) would correspond to an angular size of $\sim2$~mas.  The
angular size of an ordinary supernova, by contrast, is expected to be
$\lesssim 0.2$~mas, since supernova shells expand with velocities of
$\lesssim 20,000$~\kms\ \citep[see e.g.,][]{VLBA10th}.

A clear observational distinction between radio emission from a
relativistic GRB jet and that from a supernova shell is thus possible.
In this letter, we describe VLBI
observations of SN~2001em that we carried out in order to determine
its size, and address the question of whether the radio emission is
due to a relativistic GRB jet or a supernova.  An earlier attempt was
made by \citet{Stockdale+2005} to determine a size with VLBI, who
obtained an upper limit of $1.9 \times 0.8$~mas, corresponding to a
non-conclusive upper limit of $\sim 260,000$~\kms\ or $0.9\,c$ at
80~Mpc on one-sided expansion.\footnote{\citet{Stockdale+2005} cite a
limit on {\em two-sided} expansion of 150,000~\kms.  It is likely,
however, that the jet emission dominates the counter-jet emission, and
hence we cite the more conservative velocity limit on one-sided
expansion.  We further convert their limit, which was derived using a
distance of 90~Mpc, to our assumed distance of 80~Mpc.}

\section{OBSERVATIONS}

Because of the low flux density of only $\sim 1.5$~mJy, we used the
``High Sensitivity Array'', which consists of NRAO's\footnote{The
National Radio Astronomy Observatory, NRAO, is operated under license
by Associated Universities, Inc., under cooperative agreement with
National Science Foundation, NSF.} Very Long Baseline Array (VLBA;
each of 25~m diameter), phased Very Large Array, (VLA; 130~m
equivalent diameter), and Robert C. Byrd telescopes (GBT;
$\sim$105~m diameter), in addition to the Arecibo Radio Telescope\footnote{The
Arecibo Observatory is part of the National Astronomy and Ionosphere
Center, which is operated by Cornell University under a cooperative
agreement with the NSF.} (Ar; 305~m diameter) and the Effelsberg Radio
Telescope\footnote{The 100~m telescope at Effelsberg is operated by
the Max-Planck-Institut f\"{u}r Radioastronomie in Bonn, Germany.}
(Eb; 100~m diameter).  We phase-referenced our VLBI observations to
obtain both reliable starting models for possible self-calibration and
accurate astrometric positions. Our primary phase-reference source was
\objectname{J2145+11}, which is 1.4\arcdeg\ away on the sky.  However,
we spent two periods of $\sim$20~min phase-referencing to J2139+14,
(2.1\arcdeg\ away) which is a ``defining source'' of the International
Celestial Reference Frame \citep[ICRF;][]{Ma+1998} in order to
determine coordinates for SN~2001em tied to the ICRF.

The VLBI observations were carried out on 2004 November 22, with a
total time of 12~hrs.  We used a frequency of 8.4~GHz with a bandwidth
of 32~MHz, and we recorded both senses of circular polarization.  As
usual, a hydrogen maser was used as a time and frequency standard at
each telescope. The data were recorded with the VLBA or the MKIV VLBI
systems with a sampling rate of 256~Mbits~s$^{-1}$, and correlated
with the NRAO VLBA processor at Socorro.  The analysis was carried out
with NRAO's Astronomical Image Processing System (AIPS)\@.  The
initial flux density calibration was done through measurements of the
system temperature ($T_{\rm sys}$ at each telescope, and then improved
through self-calibration of the phase-reference sources
\objectname{J2145+11} and \objectname{J2139+14}.  We used a cycle time
of $\sim$250~s, with $\sim$170~s spent on SN~2001em.  The data from
the Kitt Peak VLBA station were of very poor quality and not used in
the final analysis.  Finally, in addition to using the phased VLA,
which was in the A configuration, as an element in our VLBI array, we
also used it as a stand-alone interferometer to determine the
integrated flux density of SN~2001em.

\section{RESULTS}

Using the VLA as a stand-alone interferometer, we found the integrated
flux density of SN~2001em on 2004 November 22 to be $1.50 \pm
0.10$~mJy at 8.4~GHz, consistent with the earlier value of $1.48 \pm
0.05$~mJy obtained on 2004 January by \citet{Stockdale+2004}, and
suggesting that the flux density at 8.4~GHz is still increasing or at
least staying relatively constant.

In Figure 1 we show the VLBI image of SN~2001em.  No
structure is apparent, and the source is at best marginally resolved
at our resolution of $1.3 \times 0.6$~mas at p.a.\ $-4$\arcdeg\ (FWHM
of a Gaussian; p.a.\ is east of north).  In particular, no structure
larger than $1.3$~mas or $1.6\ex{18}$~cm is visible.
\citet{Taylor+2004} found a radio component in GRB~030329 which, if it
is related to GRB~030329, suggests an apparent expansion velocity of
$\sim$19$\;c$.  We made large images of 0.2\arcsec\ $\times$
0.2\arcsec, to search for such components, but found none: the
brightest points except for SN~2001em were no brighter than expected
from the noise, or $<0.1$~mJy/bm.
In addition, our phase-referenced VLBI observations give an accurate
position for SN~2001em of R.A. = \Ra{21}{42}{23}{60936(2)}, decl.\ =
\dec{12}{29}{50}{3003(3)} (J2000; the figure in brackets gives the
estimated uncertainty in the last
digit).\footnote{This position is that of the radio brightness
peak. It was determined with respect to the phase calibrator source
J2139+14, for which we assumed the ICRF position of R.A. =
\Ra{21}{39}{1}{30927(2)}, decl.\ = \dec{14}{23}{35}{9920(3)}
\citep[J2000;][]{Ma+1998}.  Our estimated uncertainties are those from
our phase-referencing analysis (150~\muas), added in quadrature to the
listed uncertainties of the ICRF position for J2139+14
($\sim$300~\muas).  We note also that the differential position to our
other reference source, J2145+11, is somewhat more accurate due to the
higher signal-to-noise ratio and the somewhat smaller angular distance
to SN~2001em.  However, since the uncertainty of J2145+11's position
is 10~mas, the absolute position derived from it is less accurate.}

\begin{figure}
\label{fvlbiimg}
\epsscale{1.2}
\plotone{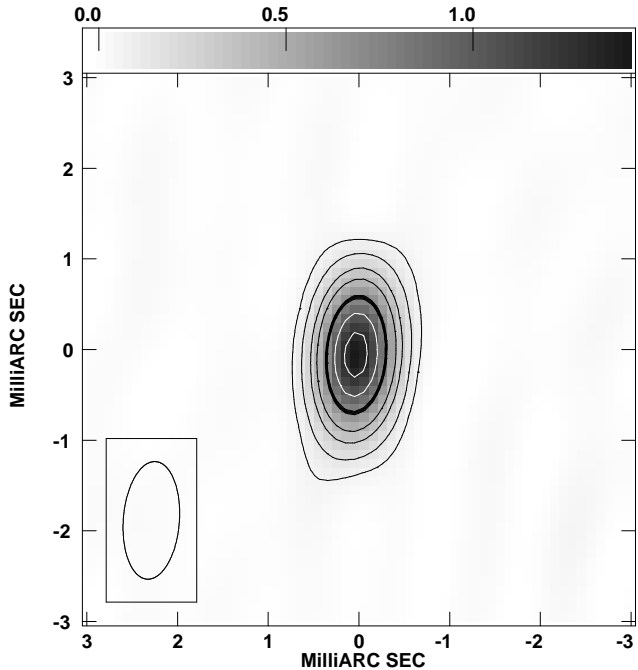}
\caption{
A VLBI image of SN~2001em on 2004 November 22.  The peak brightness
was 1.46 m\Jb\ and the rms background was 17~$\mu$\Jb.  The contours
are drawn at $-4$, 4, 10, 20, 30, {\bf 50}, 70 and 90\% of the peak
brightness, with the lowest contour being at $3\sigma$, and the 50\%
contour being emphasized. The FWHM size of the Gaussian restoring
beam, which can be compared to the 50\% contour of the source, was
$1.30 \times 0.62$~mas at p.a.\ -4\arcdeg, and is indicated at lower
left.  The greyscale is labeled in m\Jb. The origin is at R.A. =
\Ra{21}{42}{23}{60936}, decl.\ = \dec{12}{29}{50}{3003} (J2000).
North is up and east  to the left.}
\end{figure}

To get a more accurate determination of the size, we turned to model
fitting.  This process involves the least-squares fitting of
geometrical models to the VLBI visibility data in the Fourier
transform plane, and is described in more detail in, e.g.,
\citet{SN93J-P1} and \citet{SN93J-P2}.  It has the advantage of giving
results independent of the convolving beam.

Since the source is at best marginally resolved, we must choose the
model {\em a priori}. We use two different models, the first being an
elliptical Gaussian, which we choose to model a possibly elongated
source such as a jet, and the second being a spherical shell model
appropriate for a supernova.  In particular, our shell model consists
of the projection of a spherically symmetrical, optically thin shell
of emission, with the shell width being 20\% of the outer radius.
Such a shell model is
suggested by observations of SN~1993J \citep[][see also Marcaide et
al.\ 1995, who find a somewhat thicker shell]{SN93J-P1, SN93J-P2,
SN93J-P3}.

The statistical uncertainties can be determined in the fitting
process, since those of the visibility measurements are known.  The
visibility data, however, may suffer from additional uncertainties due
to possible calibration errors, which are not statistically
independent.  We thus use the square root of the visibility weights in
our fits, which results in a more robust fit less dominated by a few
very sensitive baselines.  The fit size is probably most sensitive to
the amplitude gain calibration, since for a marginally resolved source
like SN~2001em, the fit source size is correlated with the individual
antenna amplitude gains.  Our antenna amplitude gains were derived
first from the $T_{\rm sys}$ measurements at each telescope and then
refined by self-calibration using the reference sources.  They should
be un-biased, since they were derived without reference to SN~2001em
itself, and accurate to $<5$\%.  Our conservative uncertainties are
obtained by letting the antenna amplitude gains be free parameters in
the fit, which allows for possible correlations between them and the
fit size.  Our best-fit size, however, is the unbiased one arrived at
by fixing the antenna gains at the values derived from the reference
sources. This best-fit size is consistent with that derived by
freeing the antenna gains.

Using an elliptical Gaussian model appropriate for a jet, we obtained
a FWHM major axis of $0.23_{-0.18}^{+0.12}$~mas; the $3\sigma$ upper
limit on the FWHM major axis size is thus 0.59~mas.  The nominal p.a.\
of the major axis was $\sim 97$\arcdeg; however we find that neither
the ellipticity nor the p.a.\ is reliably determined by our data.

For the spherical shell model appropriate for a supernova, we find an
outer angular radius of $0.17_{-0.10}^{+0.06}$~mas, with a
corresponding $3\sigma$ upper limit of 0.35~mas.  We note that other
models with a geometry different from a spherical shell will give
different outer radii, but for likely geometries, the differences are
$<25$\%. For example, a uniform sphere gives a radius $\sim 16$\%
larger, whereas a very thin shell gives a radius $\sim 10$\% smaller
\citep[see e.g.,][]{SN93J-P2, Marscher1985}.

\section{DISCUSSION}

We have made a VLBI image of SN~2001em, a supernova with an unusually
delayed onset of radio emission, which was thought possibly to be due
to a relativistic jet not aligned with the line of sight. Such a jet
would link SN~2001em to a GRB\@.  The radio morphology of a
relativistic jet would be expected to be elongated.
\citet{GranotL2003} showed that the radio emission from such a jet
would peak near the time when the jet expansion becomes
non-relativistic.  Thus up to the time of the radio peak, a transverse
jet should be seen to expand at an apparent speed of $\sim c$, with
the apparent expansion speed being superluminal for a jet directed
toward us, and sub-luminal for a jet directed away.  Since the radio
flux density has not yet started to decline significantly, we would
expect such a jet in SN~2001em to still be in its relativistic
expansion phase, and thus the brighter jet directed toward us should
expand with an apparent speed of at least $c$. 
In the case that the jet inclination angle is near 90\arcdeg, we would
expect the jet and counterjet to be of nearly equal radio brightness,
and an end-to-end apparent expansion velocity of $\sim 2\,c$.  At 80~Mpc, this
implies a current angular size of $\gtrsim 2$~mas for the approaching
jet.

At epoch 2004.9, we found a $3\sigma$ upper limit to the FWHM major
axis size of SN~2001em of 0.59~mas, corresponding to a size of
$<7.1\ex{17}$~cm.  Assuming one-sided expansion, this corresponds to a
projected average velocity since the explosion of $<70,000$~\kms\ or
$<0.23\,c$.  In the case of a two-sided jet, the velocity of each jet
would be $<35,000$~\kms.
Since no extended emission is visible in our image, any source
expanding with a projected velocity of $> 0.5\, c$ can be responsible
for at most 10\% of the radio flux density.

This upper limit on the expansion velocity implies that the bulk of
the radio emission cannot be due to a relativistic jet associated with
SN~2001em. We note that a relativistic jet aligned near the line of
sight is unlikely, as it would be seen to expand super-luminally.  If
there is a radio-bright jet, it must have ceased relativistic
expansion at an age of $\lesssim 0.5$~yr.  We cannot of course exclude
the possibility of the body of the jet being faint in the radio, with
either the core or a terminal hot-spot being responsible for the bulk
of the radio emission.  We note that future astrometric measurements,
using the same reference sources we used above, could allow a
determination of the proper motion. In the case of a bright terminal
hot-spot moving at $\gtrsim 40,000$~\kms, the proper motion would be
$\gtrsim 0.1$~mas~yr$^{-1}$, which is easily detectable in a few
years.

The upper limit on the expansion velocity of SN~2001em is, on the
other hand, consistent with those usually seen in radio supernovae, in
which the radio emission originates from the interaction between
the supernova shell, which expands with velocities of order
10,000~\kms, and the circumstellar medium.  If we assume a spherical
shell geometry for SN~2001em, we find its outer angular radius to be
$0.17_{-0.10}^{+0.06}$~mas, corresponding to a radius of
$2.0_{-1.2}^{+0.7} \ex{17}$~cm, and an expansion velocity of
$20,000_{-12,000}^{+7,000}$~\kms\ (as above this value represents the
average expansion velocity since the supernova explosion).  This
expansion velocity is consistent with those seen in other radio
supernovae \citep[e.g.,][]{VLBA10th}.

If the radio emission of SN~2001em is due to an expanding supernova
shell, then we are left with the puzzle of its delayed onset, and also
of the spectral index which is somewhat flatter than usually observed
in supernova shell emission.  The spectral index is, however,
consistent with what is observed for pulsar nebulae.  It is perhaps
possible then, that the radio emission in SN~2001em is due to the
turn-on of a pulsar nebula.  As such it would be $\sim 5000$ times as
radio-luminous as the 1000-year old Crab Nebula is at present.  Such a
high luminosity is not excluded for a young pulsar nebula
\citep{BandieraPS1984, Chevalier2005}. We note that a new
component with an inverted radio spectrum was recently seen in
SN~1986J, almost 20~yr after the explosion \citep{SN86J-Sci,
SN86J-P1}.  This component, which is 200 times as radio-luminous as
the Crab Nebula, may be a young pulsar nebula.  Future monitoring of
SN~2001em will be necessary to determine whether the radio emission is
due to a supernova shell, a pulsar nebula, or something else.

\acknowledgements 

We thank M. Rupen for useful comments on the manuscript.
Research at York University was partly supported by NSERC\@.

\end{document}